\documentclass[twocolumn]{aastex61}

\newcommand{\Rnum}[1]{\uppercase\expandafter{\romannumeral #1\relax}}

\submitjournal{PASP}

\shorttitle{BL Cam}
\shortauthors{P. Zong et al.}

\begin{document}

\title{Pulsations of the SX Phe Star BL Camelopardalis}

\correspondingauthor{Ali. Esamdin}
\email{aliyi@xao.ac.cn}

\author{Peng. Zong}
\affiliation{Xinjiang Astronomical Observatory, Chinese Academy of Sciences, Urumqi, Xinjiang 830011, People's Republic of China}
\affiliation{University of Chinese Academy of Sciences, Beijing 100049, People's Republic of China}

\author{Ali. Esamdin}
\affiliation{Xinjiang Astronomical Observatory, Chinese Academy of Sciences, Urumqi, Xinjiang 830011, People's Republic of China}
\affiliation{University of Chinese Academy of Sciences, Beijing 100049, People's Republic of China}

\author{Jian Ning. Fu}
\affiliation{Department of Astronomy, Beijing Normal University, Beijing 100875, People's Republic of China}

\author{Hu Biao. Niu}
\affiliation{Xinjiang Astronomical Observatory, Chinese Academy of Sciences, Urumqi, Xinjiang 830011, People's Republic of China}
\affiliation{Department of Astronomy, Beijing Normal University, Beijing 100875, People's Republic of China}

\author{Guo Jie. Feng}
\affiliation{Xinjiang Astronomical Observatory, Chinese Academy of Sciences, Urumqi, Xinjiang 830011, People's Republic of China}
\affiliation{University of Chinese Academy of Sciences, Beijing 100049, People's Republic of China}

\author{Tao Zhi. Yang}
\affiliation{Xinjiang Astronomical Observatory, Chinese Academy of Sciences, Urumqi, Xinjiang 830011, People's Republic of China}

\author{Chun Hai. Bai}
\affiliation{Xinjiang Astronomical Observatory, Chinese Academy of Sciences, Urumqi, Xinjiang 830011, People's Republic of China}

\author{Yu. Zhang}
\affiliation{Xinjiang Astronomical Observatory, Chinese Academy of Sciences, Urumqi, Xinjiang 830011, People's Republic of China}

\author{Jin Zhong. Liu}
\affiliation{Xinjiang Astronomical Observatory, Chinese Academy of Sciences, Urumqi, Xinjiang 830011, People's Republic of China}



\begin{abstract}
We carried out photometric observations of the SX Phe star BL Cam in 2014, 2017 and 2018 using Nanshan 1-m telescope. In addition to the dominated frequency of 25.5790(3) cd$^{-1}$ and its two harmonics, an independent frequency of 25.247 (2) cd$^{-1}$, which is a nonradial mode frequency, was detected from the data in 2014. A total of 123 new times of light maxima were determined from our light curves in the three years, which, together with that published in the literature, were used to analyze the $O$$-$$C$ diagram. The change rate of the main period was derived as (1/P)(dP/dt) = -2.39 (8)$\times$10$^{-8}$ yr$^{-1}$, which is lower than that published in previous literature. A periodical change with a period of 14.01 (9) yr was found in the residuals of the $O$$-$$C$ curve fitting. If it was caused by the light-time effect, BL Cam should be a binary system. The mass of the companion was restricted as low as that of a brown dwarf. No evidence of the triple system suggested by previous authors was shown in our analysis.
\end{abstract}




\section{Introduction}
SX Phe stars are high-amplitude $\delta$ Scuti variables (HADS) belonging to Population II stars with high spatial motion and low metallity \citep{2006A&A...451..999F}. However, some of them are metal-rich and more like halo A-type stars found in the $Kepler$ field of view \citep{2017MNRAS.466.1290N}. The SX Phe stars have masses of $\sim$1.0 $-$ 1.2 M$_\odot$ and relatively young ages $\sim$2.0 $-$ 5.0 Gyr \citep{1990ASPC...11...64N}. They are mostly located in the blue stragglers of globular clusters \citep{2015MNRAS.449.3535M, 2015A&A...578A.128K, 2013AcA....63...91K} while some of them are also found in the field. However, there are no observation differences in the physical properties of SX Phe stars in the field and in the globular clusters \citep{2011AJ....142..110M}. The pulsating behavior of the SX Phe stars in the field is similar to the Population I high amplitude $\delta$ Scuti stars \citep{2015pust.book.....C}. Most SX Phe stars have short periods ($\lesssim$ 0.08 d) and large amplitudes ($>$ 0.3 mag) \citep{2006A&A...451..999F}. 
It is generally assumed that the metal-poor HADS show only two radial pulsation modes, but BL Cam ($\alpha_{2000}$ = 03$^{h}$47$^{min}$19$^{s}$, $\delta_{2000}$ = +63$^{o}$22$^{'}$07$^{''}$, $<V>$ $\sim$ 13.1 mag, $\Delta$ V $\sim$ 0.33 mag, \citealt{2001A&A...366..178R}) as a SX Phe star in the field seems to be an exception \citep{ 2011AJ....142..110M}. 

BL Cam was discovered by \cite{1970LowOB...7..183G} and considered as a possible candidate of white dwarf. \cite{1977ApJ...215L..25B} noted that it is a pulsating star with the period of 0.039 days and the amplitude of 0.33 mag. \cite{1997PASP..109.1221M} determined its metal abundance of [Fe/H] = $-$2.4 and classfied it as a Population II star. Its multiperiodic nature has been studied by previous authors \citep{2006A&A...451..999F, 2007A&A...471..255R, 2008ChJAA...8..237F}. 
\cite{1997PASP..109...15H} reported its first overtone of 32.679 cd$^{-1}$, and the period ratio of the first overtone to the fundamental mode of 0.783. The first overtone around 31.6 cd$^{-1}$ was also detected by previous authors \citep{1999MNRAS.308..631Z, 1999JASS...16..241K, 2001IBVS.5061....1Z, 2002IBVS.5317....1W}, leading to the period ratio of 0.810. However, the first overtone was not detected by \cite{2008ChJAA...8..237F}. \cite{2007A&A...471..255R} carried out a multi-site photometric campaign of BL Cam and detected 21 independent pulsation frequencies (excluding the fundamental mode) with amplitudes ranging from 1.6 to 7.4 mmag. The first overtone and the period ratio they obtained are the same with that of \cite{1997PASP..109...15H}. The multiperiodic nature of BL Cam and the controversies of its first overtone make it deserve further study. 

\cite{1997PASP..109...15H} discovered the long-term increasement of the main period using the $O$$-$$C$ method. \cite{1998A&A...332..958B} performed an analysis of the $O$$-$$C$ diagram, revealing an increasement of 0.02 s of the main period over the past 20 years with the period variation rate of (1/P)(dP/dt) = 2.90$\times$10$^{-7}$yr$^{-1}$. \cite{2003PASP..115..755K} pointed out a reversal trend in the $O$$-$$C$ diagram of BL Cam since 1999, which led to a possible sinusoidal behavior, suggesting that BL Cam might be a binary system. The binary system analysis of $O$$-$$C$ diagram of BL Cam has been published in previous literature \citep{2006A&A...451..999F, 2008ChJAA...8..237F}. \cite{2010A&A...515A..39F} performed a triple system analysis of the $O$$-$$C$ diagram. However, the determination of the second companion's parameters was not successful. \cite{2013PASP..125..639C} suggested that there might be an abrupt change in the main period, probably because BL Cam is a triple system. 

To study the characteristics of pulsation and behaviors of the main period of BL Cam, CCD photometric observations had been performed in 2014, 2017, and 2018. Section 2 of this paper presents observations and data reduction. Sections 3 and 4 give the results of frequency analysis and the $O$$-$$C$ diagram diagnosis, respectively. We discuss our results in Section 5. The summary of this work is given in Section 6.

\section{Observations and Data reduction} \label{sec:data}
BL Cam was observed using the 1-m Nanshan Optical Wide-field Telescope (NOWT) of Xinjiang Astronomical Observatory (XAO) in 2014, 2017, and 2018. 
The location of NOWT is $87.1^{\circ}$ E and $43.8^{\circ}$ N, near Urumqi of Xinjiang in China. The telescope was equipped with a standard Johnson multicolor filter system (i.e., U B V R I filters) (e.g. \citealt{1976MmRAS..81...25C}) and an E2V CCD 203-82 (blue) camera mounted on the primary focus of the telescope. The CCD camera, which has 4K$\times$4K pixels and a $1.3^{\circ}$ $\times$ $1.3^{\circ}$ field of view, operated at a temperature of about $-$120$^{\circ}$C cooling with liquid nitrogen \citep{2016RAA....16..154S, 2018Ap&SS.363...68M}.

The exposure time of observations was either 3 s or 5 s in 2014, 10 s in 2017, and 12 s in 2018. The exposure time increased in different observation seasons in order to get a high signal-to-noise ratio (S/N).
A total of 11,854 CCD images are obtained on 14 nights using V filter, including 6831 images on nine nights in 2014, 4173 images on four nights in 2017 and 850 images on one night in 2018. The journal of observations of BL Cam is given in Table~\ref{tb:e_t}. 

The IRAF \footnote{Image Reduction and Analysis Facility is developed and distributed by the National Optical Astronomy Observatories, which is operated by the Association of Universities for Research in Astronomy under operative agreement with the National Science Foundation.} \citep{1986SPIE..627..733T, 1993ASPC...52..173T} package was used to reduce all CCD images by subtracting the bias and making flat field correction using the master flat. The dark correction was not considered because the thermal noise of the CCD camera was negligible with the value less than 1 e pix$^{-1}$h$^{-1}$ at the temperature of $-$120$^{\circ}$C \citep{2018RAA....18....2Y}.

\begin{figure}
\centering
	\includegraphics[width=6.5cm, angle=0]{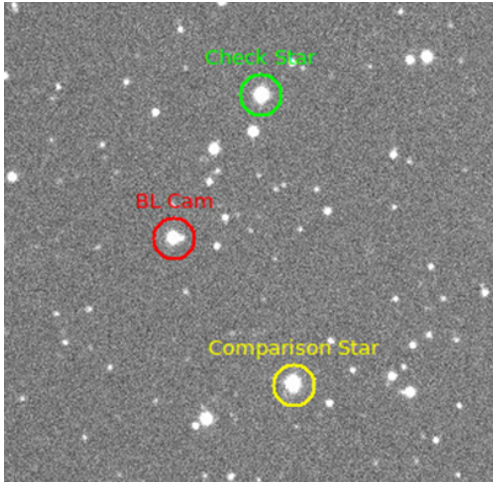}
    \caption{Trimmed CCD image of BL Cam collected by NOWT. BL Cam, the comparison and the check star are marked with red, yellow, and green circles, respectively.}
    \label{fig:object}
\end{figure}

\begin{figure*}
\centering
	\includegraphics[width=10.5cm, angle=-90]{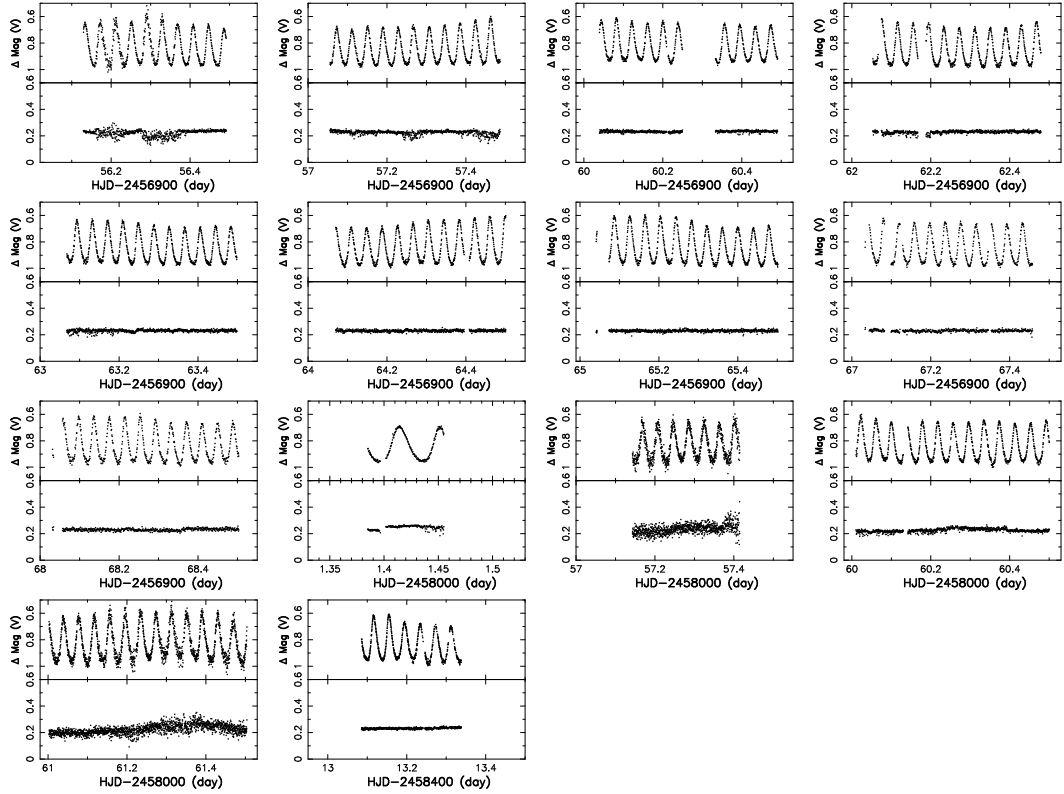}
    \caption{Light curves of BL Cam in V band in 2014, 2017, and 2018. The top panel in each subfigure show the magnitude difference between BL Cam and the comparison star, while the bottom panels present the magnitude difference between the check star and the comparison star.} 
    \label{fig:14_lg}
\end{figure*}

Two stars selected by \cite{2013PASP..125..639C} were taken as the comparison and the check star, respectively, as shown in Figure~\ref{fig:object}. The standard deviations of the differential magnitudes between the check star and the comparison star yield the estimations of the photometric precisions, with typical values of 0$^{m}$.005 and 0$^{m}$.035 in good and poor observational conditions, respectively. The light curves of BL Cam were obtained by calculating the differential magnitudes between the target and the comparison star. For frequency determination, every data point of the light curve takes its own weighting, which is the inverse square of the photometric precision. Figure~\ref{fig:14_lg} shows the light curves of BL Cam in V band in 2014, 2017 and 2018, respectively.

\begin{table*}
\centering
	\caption{Journal of observations for BL Cam in 2014, 2017 and 2018.}
\label{tb:e_t}
	\begin{tabular}{ccccc} 
		\hline
		\hline
		Date &Exposure Time &Filter &CCD Images& Length of Observation \\
		     &  (second)      &       &          &  (hour)                 \\
		\hline
		 2014.10.25&3 &V& 713&8.7 \\
	     2014.10.26&3 &V& 854 &10.4\\
		 2014.10.29&3 &V& 707 &8.6\\
	     2014.10.31&3 &V& 788 &9.6\\
	     2014.11.01&5 &V& 876 &10.7\\
	     2014.11.02&5 &V& 846 &10.3\\
		 2014.11.03&5 &V& 874 &10.6 \\
	     2014.11.05&5 &V& 564 &6.9 \\
	     2014.11.06&5 &V& 609 &7.4 \\
         2017.09.04&10 &V&216 &2.6\\
	     2017.10.30&10 &V&1007&6.5 \\
		 2017.11.02&10 &V&1089&7.0\\	     
	     2017.11.03&10 &V&1861&12.0\\
	     2018.10.21&12 &V& 850& 6.0\\
		 \hline
		 total&    &     & 11854&117.3\\
		 \hline
	\end{tabular}
\end{table*}

\section{Frequency Analysis}
Frequency analysis was performed with the light curves in 2014 by using Period04 \citep{2005CoAst.146...53L}, which focuses on multi-frequency analysis using Fourier transform and least-square fitting. The following formula was applied to the fitting of the light curves:
\begin{equation}
m = m_0+\sum A_i sin[2 \pi (f_i t + \varphi_i)]
\label{eq:zp_H}
\end{equation}
where $A_i$ , $f_i$ and $\varphi_i$ are the amplitude, frequency, and phase, respectively. We follow a standard pre-whitening procedure. 

First, the highest peak in the spectrum of the light curves was selected as a significant frequency, which was taken to do least-square fitting to determine the 
values of frequency, amplitude, and phase. Theoretical light curves were then constructed with the solution of the determined frequency and subtracted from the origin light curves to get the residual light curves. Then, we picked out the highest peak in the spectrum of the residual light curve as the next significant frequency and subtracted the light curves constructed with the solutions of all the detected frequencies from the origin light curves. We repeated the above steps until no significant frequency was detected. We adopted S/N $>$ 6 \citep{2015MNRAS.448L..16B, 2015A&A...579A.133B} as the criterion of resolving the frequencies. Table~\ref{tb:zp_14} lists the significant frequencies detected from the light curves in 2014.

\begin{figure*}
\centering
	\includegraphics[width=7.0cm, angle=-90]{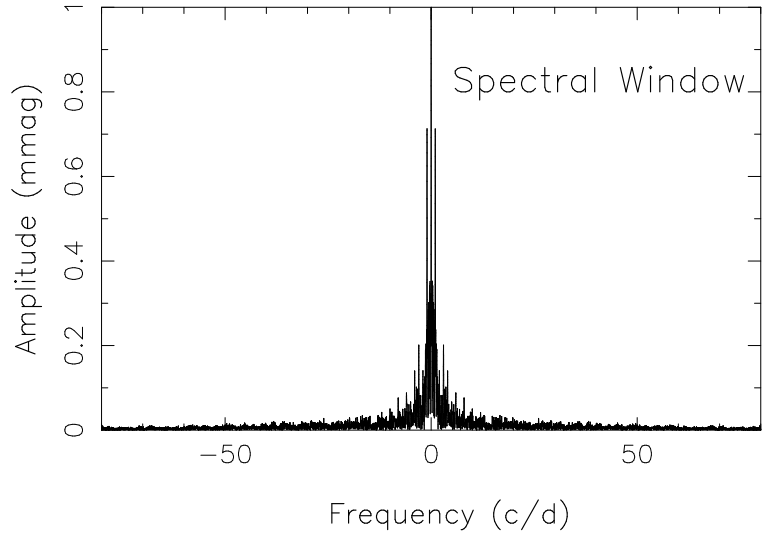}
    \caption{Spectral window of the light curves of BL Cam in V band in 2014.}
    \label{fig:14_1}
\end{figure*}

\begin{figure*}
\centering
	\includegraphics[width=11.5cm, angle=-90]{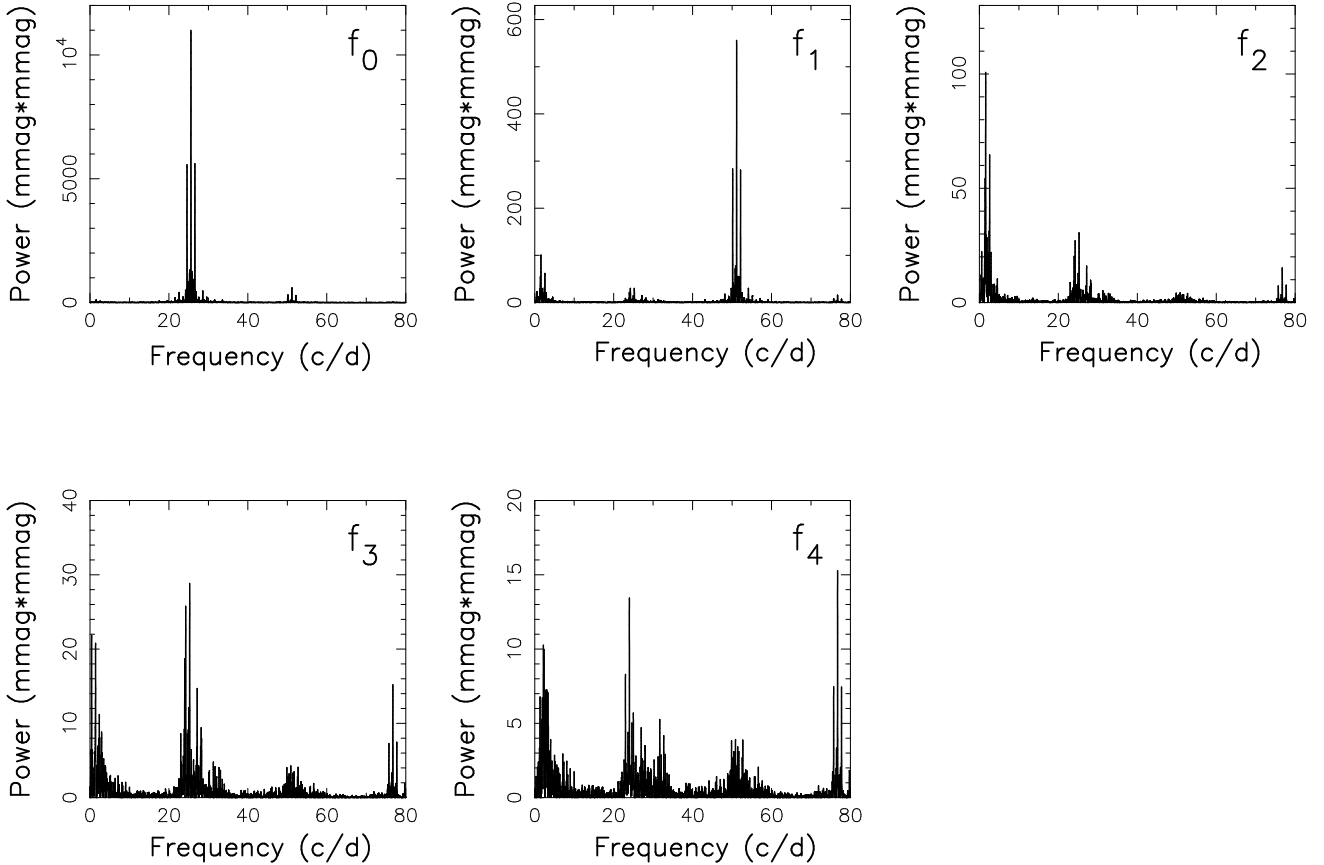}
    \caption{Fourier power spectra of the frequency pre-whitening process for the light curves of BL Cam in V band in 2014.}
    \label{fig:14}
\end{figure*}

The dominated frequency f$_0$ = 25.57827 (9) cd$^{-1}$ and its two harmonics f$_1$ (2f$_0$), f$_4$ (3f$_0$) were detected from the light curves in 2014. In Table~\ref{tb:zp_14}, the frequency detected in the region of 0$-$3 cd$^{-1}$ was not considered to be significant frequencies of the variable star, as it may very possibly be caused by either the instrument sensitivity instability or variations of the atmospheric transparency as suggested by \cite{2012AJ....144...92Y} and \cite{2018RAA....18....2Y}. The frequency of f$_3$, which is not linked in any way with the other significant frequencies, is considered as an independent frequency. Figures~\ref{fig:14_1} and \ref{fig:14} show the spectral window and Fourier power spectra of frequency pre-whitening process for the light curves of BL Cam in 2014, respectively.

Because the data collected in 2017 and 2018 covered only four nights and one night respectively, we did not try to do frequency analysis with the data in these two years.

\begin{table*}
\centering
	\caption{Frequencies Detected from the Light Curves in 2014. }
	\label{tb:zp_14}
	\begin{tabular}{lllll} 
		\hline
		\hline
		ID&Freq.& Ampl.& S/N& Identification\\ 
		  &  (cd$^{-1}$)&  (mmag)     &    &                \\
		\hline
f$_0$&	25.57827 (9)&	 148 (1) &	141.0 &Main frequency \\
f$_1$&	51.1566 (4)&	 33.3 (7) &	 46.0 & 2f$_0$\\
f$_2$&  1.5840 (7) & 17.0 (2)& 7.2 &  Alias \\
f$_3$&	25.247 (2)&	 8.0 (1) &	 7.5 & Nonradial\\
f$_4$&	76.731 (2)&	 5.7 (4) &	16.1 & 3f$_0$\\

\hline
	\end{tabular}
\end{table*}

\section{Period Variation Analysis}
\subsection{Times of maximum light}
The $O$$-$$C$ method is a classical method for studying the periodic variations of the variable star as it is sensitive to the cumulative effect of periodic changes \citep{2005ASPC..335....3S,2007C&T...123R.122D}. The times of maximum light of BL Cam were derived by fitting a third or fourth order polynomial around each peak of the light curves, the median uncertainty is $\sigma$ = 0.0001 d. In total, 123 new times of light maxima were determined and listed in Table~\ref{tb:zp_O}. Together with the 1460 times of light maxima published in the previous literature \citep{1977ApJ...215L..25B,1978PASP...90..275M,1990IBVS.3428....1R,1997PASP..109...15H,1999MNRAS.308..631Z,2000JRASC..94..124B,2000IBVS.4872....1V,2001IBVS.5061....1Z,2002IBVS.5317....1W,2003PASP..115..755K,2006A&A...451..999F,2008ChJAA...8..237F,2010A&A...515A..39F,2013PASP..125..639C}, we tried to reanalyze the $O$$-$$C$ diagram of BL Cam. 

\begin{table*}
  \centering
   \caption{Times of Light Maxima of BL Cam.}
	\label{tb:zp_O}
	\begin{tabular}{cccccc}
\hline
\hline
 Maxima   &  Cycle  &  $O$$-$$C$&  Maxima &  Cycle & $O$$-$$C$ \\
 \hline 
 HJD    &   E   & day&HJD  &   E  &  day\\
 \hline
2456956.13366&353736&0.00033&	2456965.20376&353968&-0.00029\\
2456956.17328&353737&0.00085&	2456965.24336&353969&0.00022\\
2456956.21215&353738&0.00062&	2456965.32123&353971&-0.00011\\
2456956.25144&353739&0.00082&	2456965.36053&353972&0.00009\\
2456956.29045&353740&0.00073&	2456965.39965&353973&0.00011\\
2456956.33207&353741&0.00325&	2456965.43940&353974&0.00077\\
2456956.36826&353742&0.00034&	2456965.47771&353975&-0.00002\\
2456956.40786&353743&0.00084&	2456967.15864&354018&-0.00030\\
2456956.44660&353744&0.00049&	2456967.19770&354019&-0.00034\\
2456956.48530&353745&0.00009&	2456967.23619&354020&-0.00095\\
2456957.07249&353760&0.00081&	2456967.27583&354021&-0.00041\\
2456957.11114&353761&0.00036&	2456967.31501&354022&-0.00032\\
2456957.15075&353762&0.00087&	2456967.39259&354024&-0.00094\\
2456957.18986&353763&0.00089&	2456967.43215&354025&-0.00048\\
2456957.22806&353764&-0.00001&	2456967.47089&354026&-0.00083\\
2456957.26659&353765&-0.00058&	2456968.09761&354042&0.00032\\
2456957.30750&353766&0.00123&	2456968.13594&354043&-0.00045\\
2456957.34717&353767&0.00180&	2456968.17561&354044&0.00012\\
2456957.38516&353768&0.00070&	2456968.21462&354045&0.00003\\
2456957.42406&353769&0.00050&	2456968.25295&354046&-0.00073\\
2456957.46288&353770&0.00022&	2456968.29210&354047&-0.00068\\
2456960.04380&353836&0.00068&	2456968.33114&354048&-0.00074\\
2456960.08201&353837&-0.00021&	2456968.37107&354049&0.00009\\
2456960.12129&353838&-0.00003&	2456968.41068&354050&0.00061\\
2456960.16170&353839&0.00129&	2456968.44969&354051&0.00052\\
2456960.19956&353840&0.00005&	2456968.48851&354052&0.00024\\
2456960.23783&353841&-0.00078&	2458001.41417&380471&-0.00188\\
2456960.35630&353844&0.00040&	2458001.45229&380472&-0.00286\\
2456960.39474&353845&-0.00026&	2458057.16948&381897&-0.00019\\
2456960.43451&353846&0.00041&	2458057.20663&381898&-0.00214\\
2456960.47344&353847&0.00024&	2458057.24689&381899&-0.00098\\
\hline
\end{tabular}
 \end{table*}

\begin{table*}
  \centering
    
	\begin{tabular}{cccccc}
\hline
\hline
2456962.11588&353889&0.00057&	2458057.28520&381900&-0.00177\\
2456962.15526&353890&0.00085&	2458057.32509&381901&-0.00097\\
2456962.23310&353892&0.00050&	2458057.36264&381902&-0.00252\\
2456962.27196&353893&0.00026&	2458057.40222&381903&-0.00204\\
2456962.31194&353894&0.00114&	2458060.02277&381970&-0.00105\\
2456962.35191&353895&0.00201&	2458060.09999&381972&-0.00203\\
2456962.39092&353896&0.00192&	2458060.17873&381974&-0.00148\\
2456962.42940&353897&0.00131&	2458060.21731&381975&-0.00200\\
2456962.46797&353898&0.00078&	2458060.25886&381976&0.00045\\
2456963.09316&353914&0.00040&	2458060.29577&381977&-0.00173\\
2456963.13264&353915&0.00078&	2458060.33383&381978&-0.00277\\
2456963.17104&353916&0.00009&	2458060.37357&381979&-0.00213\\
2456963.20961&353917&-0.00044&	2458060.41272&381980&-0.00208\\
2456963.24855&353918&-0.00060&	2458060.45320&381981&-0.00070\\
2456963.32673&353920&-0.00062&	2458061.03836&381996&-0.00201\\
2456963.36631&353921&-0.00013&	2458061.07681&381997&-0.00265\\
2456963.40563&353922&0.00009&	2458061.11725&381998&-0.00131\\
2456963.44405&353923&-0.00059&	2458061.15632&381999&-0.00134\\
2456963.48333&353924&-0.00041&	2458061.19537&382000&-0.00139\\
2456964.10916&353940&-0.00014&	2458061.23450&382001&-0.00135\\
2456964.14821&353941&-0.00019&	2458061.27388&382002&-0.00107\\
2456964.18739&353942&-0.00011&	2458061.31669&382003&0.00264\\
2456964.22727&353943&0.00067&	2458061.39057&382005&-0.00168\\
2456964.30498&353945&0.00019&	2458061.47106&382007&0.00062\\
2456964.34423&353946&0.00034&	2458413.11605&391001&-0.00103\\
2456964.38365&353947&0.00066&	2458413.15456&391002&-0.00161\\
2456964.42206&353948&-0.00003&	2458413.19435&391003&-0.00092\\
2456964.46020&353949&-0.00099&	2458413.23428&391004&-0.00009\\
2456965.08672&353965&-0.00003&	2458413.27298&391005&-0.00049\\
2456965.12622&353966&0.00037&	2458413.31186&391006&-0.00071\\
2456965.16517&353967&0.00022&	             &      &        \\
\hline
\end{tabular}
 \end{table*}

\subsection{$O$$-$$C$ diagram}
Using the 1583 times of the light maxima and the linear ephemeris given by \cite{1999MNRAS.308..631Z}:
\begin{equation}
HJD = 2443125.8015 + 0.03909785 E,
\label{eq:H_H}
\end{equation}
the number of cycle (E) for each observed maximum is calculated. Linear fitting of Heliocentric Julian Day (HJD) and cycle number (E) yields a new period value of 0.0390979135 (9) and a new linear ephemeris as
\begin{equation}
HJD = 2443125.7938 (4) + 0.0390979135 (9) E,
\label{eq:zp_H}
\end{equation}
Using the new linear ephemeris (Equation~\ref{eq:zp_H}), the values of $O$-$C$
could be determined. The times of light maxima determined in this work, the numbers of cycle (E) and the values of $O$-$C$ are listed in Table~\ref{tb:zp_O}. 
\cite{2013PASP..125..639C} show that a discontinues change in the main period is supported by the whole data of $O$$-$$C$ diagram. In this paper, we do not intend to explore the discontinue change of the main period, which was discussed by them. \cite{2008ChJAA...8..237F} carried out the binary analysis for the $O$$-$$C$ diagram on the data of HJD $>$ 2448800 (E $>$ 150000). We do the fitting to the data of E $>$ 150000 to determine whether BL Cam is in a binary system. To determine the change rate of the main period (1/P)(dP/dt), a parabolic fit is performed for the $O$$-$$C$ diagram. Our fit is concentrated on the data with E $>$ 150000, which leads to a new quadratic ephemeris:
\begin{equation}
O-C = -2.00 (7)\times 10^{-13} \times E^2 + 1.25 (4)\times 10^{-7} \times E -0.0190 (5),
\label{eq:zp_fit}
\end{equation}
The standard deviation $\sigma$ of the residuals of the quadratic fitting is 0.0015 days. The corresponding period change rate is (1/P)(dP/dt) = -2.39 (8)$\times$10$^{-8}$ yr$^{-1}$.

\begin{figure}
\centering
	\includegraphics[width=5.3cm, angle=-90]{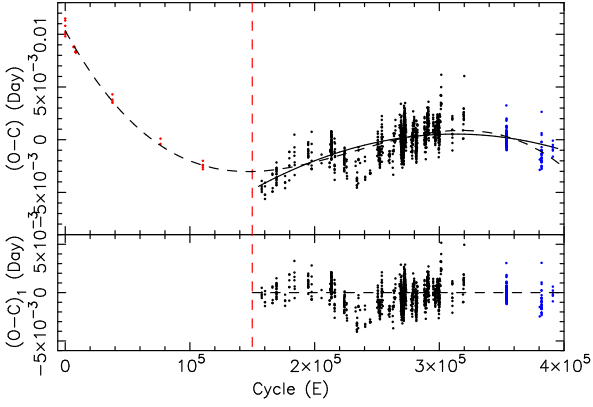}
    \caption{The $O$$-$$C$ diagram of BL Cam. The top panel shows the $O$$-$$C$ of BL Cam fitted by a quadratic ephemeris for those points in the region of E $>$ 150000, and the bottom panel shows the residuals of the quadratic fit. The blue dots are the values of $O$$-$$C$ we determined from the light curves in 2014, 2017, and 2018. The red solid dots are in the region of E $<$ 150000. The dotted line is the dividing line at E = 150000. The black dotted line is the cubic fitting for the data of $O$$-$$C$ diagram.}
    \label{fig:second_order}
\end{figure}

\begin{figure}
\centering
	\includegraphics[width=5.3cm, angle=-90]{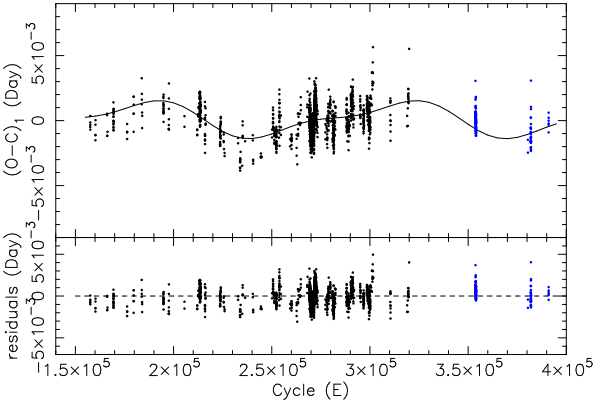}
    \caption{Residuals of the parabolic fit of the $O$$-$$C$ curve. The top panel shows the residuals of $O$$-$$C$ fitted by Equation~(\ref{eq:zp_f}) (solid line), and the bottom panel shows the residuals after subtracting solutions of Equation~(\ref{eq:zp_f}). }
    \label{fig:lt}
\end{figure}

When the parabolic curve (as shown in the top panel of Figure~\ref{fig:second_order}) is subtracted from the $O$$-$$C$ diagram, the residuals show a significant periodic change (as shown in the bottom panel of Figure~\ref{fig:second_order}), which might be caused by the light-time effect. The periodic variation of the residuals could be interpreted with a binary system hypothesis and plotted in Figure~\ref{fig:lt}. The $O$$-$$C$ residuals were fitted with the following formula \citep{1996A&AS..120...63B}:
\begin{equation}
(O-C)_1=a_0+\sum\limits_{i=1}^{2}[a_icos(i \omega E)+b_isin(i \omega E)],
\label{eq:zp_f}
\end{equation}

\begin{table}
\centering
\caption{The Solutions of Equation~(\ref{eq:zp_f}).}
	\label{tb:slou}
	\begin{tabular}{lll} 
		\hline
		\hline
		Coefficient & Value & Error($\sigma$) \\
		\hline
		$a_0$ &$1.1\times 10^{-4}$&$3.0\times 10^{-5}$\\
		$a_1$ &$-9.19\times 10^{-4}$&$7.0\times 10^{-6}$\\
		$b_1$  &$7.8\times 10^{-4}$ &$8.0\times 10^{-5}$\\
		$a_2$  &$4.6\times 10^{-4}$ & $4.0\times 10^{-5}$\\
	    $b_2$  &$1.5\times 10^{-4}$&$8.0\times 10^{-5}$\\
	  $\omega$ &$4.8\times 10^{-5}$&$3.0\times 10^{-5}$\\
    \hline
	\end{tabular}
\end{table}
The solutions of Equation~(\ref{eq:zp_f}) are listed in Table~\ref{tb:slou}.

\section{Discussion}
\subsection{Additional frequencies in the region of 25-26 cd$^{-1}$}
From Table~\ref{tb:zp_14}, one notes that in addition to the main frequency f$_0$ and its two harmonics, one frequency in the 0-3 cd$^{-1}$ region, there is an additional frequency f$_3$ = 25.247 (2) cd$^{-1}$ which has no numerical relation with the above frequencies. Hence, we think it is an independent frequency. As its value is very close to that of the main frequency f$_0$, which should be a radial mode due to its high amplitude, f$_3$ should be a nonradial mode. We should note that the frequencies in the region of 31.6-32.6 cd$^{-1}$ detected by previous authors \citep{1997PASP..109...15H,1999MNRAS.308..631Z,2003PASP..115..755K,2006A&A...451..999F,2007A&A...471..255R} were not detected in this work. \cite{2008ChJAA...8..237F} suggested that this might be due to the amplitude changes, leading the amplitude lower than the photometric limit of the observations. 

\begin{figure}
\centering
\includegraphics[width=12.0cm, angle=-90]{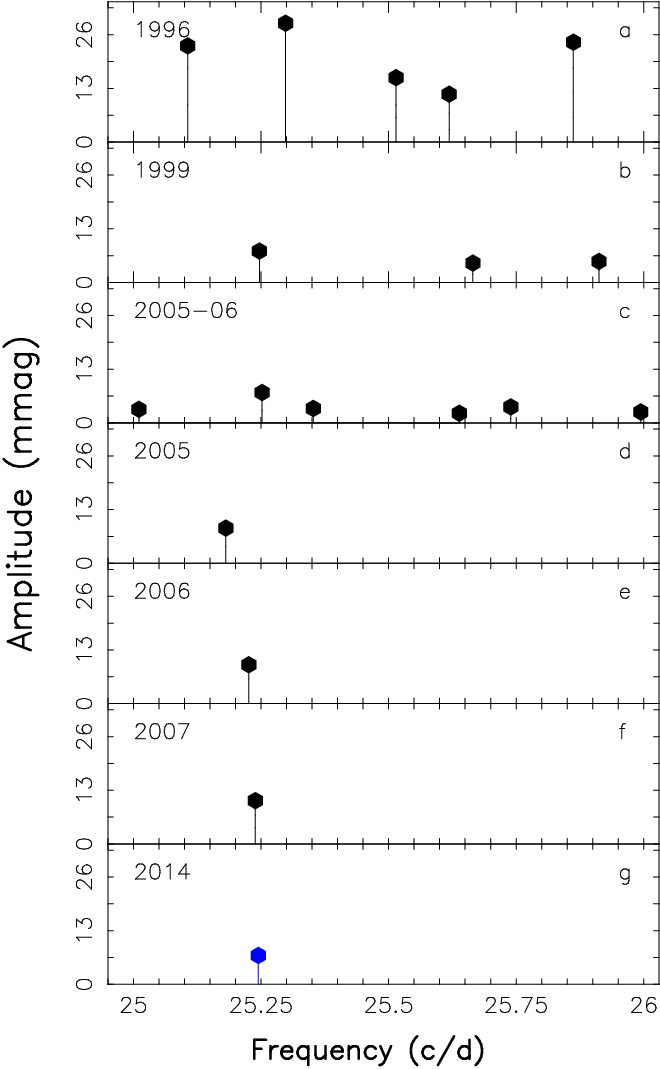}
\caption{The frequencies detected in the region of 25-26 cd$^{-1}$ in different years. a)\cite{1999MNRAS.308..631Z}, b)\cite{2001IBVS.5061....1Z}, c) \cite{2008ChJAA...8..237F}, d)\cite{2007A&A...471..255R}, e,f)\cite{2008ChJAA...8..237F} and g this work.}
\label{fig:am_25}
\end{figure}

Hence, we collect all the additional frequencies in the region of 25-26 cd$^{-1}$ of
BL Cam in the literature \citep{1999MNRAS.308..631Z, 2001IBVS.5061....1Z, 2008ChJAA...8..237F, 2007A&A...471..255R} as listed in Table~\ref{tb:zp_25}. One may find that the number of the frequency, the frequency values and the amplitudes are all different to each other. On the other hand, we see only one additional frequency in the 25-26 cd$^{-1}$ region in the bottom 4 panels of Figure~\ref{fig:am_25}. No interpretations are present to these detections.

\begin{table*}
\centering
	\caption{Frequencies detected in the region of 25-26 cd$^{-1}$ from Different Observation Data Sets. }
	\label{tb:zp_25}
	\begin{tabular}{llll} 
		\hline
		\hline
		Year& Freq.& Ampl. & Reference \\
		    & (cd$^{-1}$) &(mmag)&     \\
		\hline
		     &25.2982             &28.8&       \\
		     &25.8622             &24.2&       \\
		1996 &25.1065             &23.3&  Zhou et al.(1999)\\
		     &25.5147             &15.6&      \\
		     &25.6188             &11.6&      \\
	                                       \\
	          &25.2469            &7.6&   \\
	    1999  &25.9122            &5.1& Zhou et al.(2001)\\
	          &25.6653            &4.7&  \\
                                          \\          
	          &25.25226           &7.35& \\
	          &25.63866           &2.33&  \\
	     2005-06 &25.73943           &3.86& Rodr{\'{\i}}guez et al.(2007)\\
	          &25.01079           &3.27& \\
	          &25.35245           &3.48&   \\
	          &25.99419           &2.64&   \\
	                                       \\
	      2005&25.181             &8.5&  \\
	      2006&25.226             &9.4&  Fu et al.(2008)\\
	      2007&25.239             &10.5&   \\ 
	                                            \\
	      2014&25.247(2)          &8(1)&this work\\

	    \hline      
	\end{tabular}
\end{table*}

\subsection{Period Changes}
The change rate of the main period of BL Cam was determined by previous authors using a parabolic fit \citep{1997PASP..109...15H, 1999MNRAS.308..631Z, 2006A&A...451..999F, 2008ChJAA...8..237F, 2010A&A...515A..39F, 2013PASP..125..639C}. The change rate of the main period determined in this work is (1/P)(dP/dt) = -2.39 (8)$\times$10$^{-8}$ yr$^{-1}$. A binary system analysis on the residuals of the $O$$-$$C$ is performed in this work. The following parameters of the companion: eccentricity (e$^{'}$), orbital period (P$^{'}$), projection of the orbit radius ($A^{'}sin i^{'}$), and argument of periastron ($\omega^{'}$) listed in Table~\ref{tb:par} are determined with the formula adopted by \cite{1996A&AS..120...63B}. The values of these parameters obtained in this work are consistent with those given by the previous authors \citep{2006A&A...451..999F,2008ChJAA...8..237F} except the orbital period and the argument of periastron, which is probably because the newly determined data change the trend of residuals of the $O$$-$$C$ fit.

\cite{2008ChJAA...8..237F} suggested that the main period of BL Cam might be undergone an abrupt change, because they divided the $O$$-$$C$ data into two parts. 
A triple system analysis for BL Cam was performed by \cite{2010A&A...515A..39F}, but determining the orbital parameters of the second companion was unsuccessful. They pointed out that those significant change in $O$$-$$C$ might be due to abrupt change of the main period. \cite{2013PASP..125..639C} obtained the value of the abrupt change of the main period $\Delta$P = $-$ 0.126 (4) s by dividing $O$$-$$C$ data into two parts, i.e. the oldest data of E $<$ 150000 and the data in the region of E $>$ 150000. They suggested that the abrupt change of the main period might be caused by a third body. They also pointed out that a cubic more appropriately fits the behavior of the $O$$-$$C$ diagram as shown in top panel of Figure~\ref{fig:second_order}. But the physical meaning of the third order term is not understood. \cite{2005ASPC..335....3S} suggested that statistical validity must be considered very carefully in studying sudden changes of period. Figure~\ref{fig:second_order} shows the whole fitting of the $O$$-$$C$ diagram. The data are not sufficient to prove that the abrupt change of the main period was caused by the third object, and its orbital properties are dependent on the long-term correction of the $O$$-$$C$ trend, which need to be confirmed by future observations. 

\begin{figure}
\centering
	\includegraphics[width=5.3cm, angle=-90]{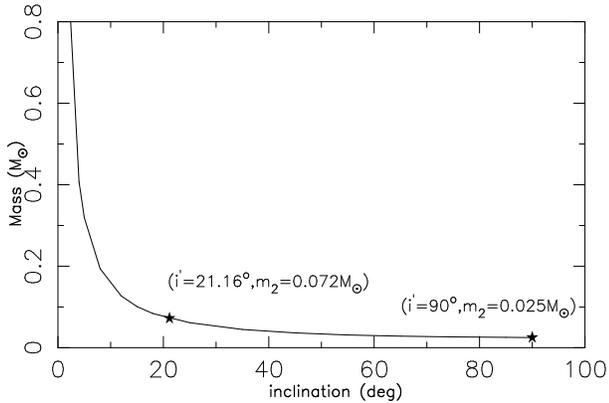}
    \caption{Inclination versus mass of the second body. The mass of the companion between the two pentagrams corresponds to the mass of brown dwarf.}
    \label{fig:mass}
\end{figure}

To determine the mass of the companion, we use the mass function adopted in previous literature \citep{2008ChJAA...8..237F, 2010AJ....139.2639L}. The value of mass function of the companion is derived to be $1.65 (3) \times 10^{-4}$ and it is 17$\%$ of that obtained by \cite{2008ChJAA...8..237F}. This might be due to the orbital period we determined in this work is different from theirs. During the observations of BL Cam we did not detect any eclipsing, hence the inclination of the system can not be precisely determined. Based on the mass of 0.99 M$_\odot$ of BL Cam determined by \cite{1997PASP..109.1221M} and the value of mass function given in this work, a relationship between the mass of the companion and the orbital inclination of binary, as shown in Figure~\ref{fig:mass}, could be determined. When the orbital inclination ranges from $21.16^{\circ}$ to $90^{\circ}$, the mass of companion is between 0.072 M$_\odot$ and 0.025 M$_\odot$. If the inclination distributes randomly, the possibility that the companion is a brown dwarf is 76$\%$.
\begin{table}
\centering
\caption{Parameters of the Companion Star of BL Cam.}
	\label{tb:par}
	\begin{tabular}{lll} 
		\hline
		\hline
		Parameter & Value & Error($\sigma$) \\
		\hline
	    $e^{'}$  &0.80&(7)\\
		$\omega^{'}$(deg) &81.29&(11)\\
		$A^{'}sin i^{'}$(AU)&0.210&(8)\\
		P$^{'}$(yr)&14.01&(9)\\
	\hline
	\end{tabular}
\end{table}

\section{Summary}
Based on new photometric observations of BL Cam in 2014, we analyzed its pulsation characteristics. An independent frequency of 25.247 (2) cd$^{-1}$ was detected in the region of 25-26 cd$^{-1}$ from the data of 2014. Using newly determined 123 times of light maxima and those published in the previous literature, the $O$$-$$C$ analysis for BL Cam reveals a period of 0.0390979135 (9) days and an update period change rate of the main period $(1/P)(dP/dt) = -2.39 (8)\times 10^{-8} yr^{-1} $. The residuals of fitting the $O$$-$$C$ curve implies that BL Cam might be a binary system in an eccentric orbit with a period of 14.01 (9) yr. The companion might be a brown dwarf. More photometric and spectroscopic observations are needed to reveal the pulsational characteristics and confirm the binary hypothesis of BL Cam. 

\acknowledgements
This research is supported by Nanshan 1 m telescope of Xinjiang Astronomical Observatory, light in China's Western Region: 2015-XBQN-A-02, XBBS-2014-25; National 
Natural of Science Fundation of China: 11873081, 11673003, 11833002, 11661161016; 2017 Heaven Lake Hundred-Talent Program of Xinjiang Uygur Autonomous Region of China.

\software{
          Period04 \citep{2005CoAst.146...53L},
		  IRAF \citep{1986SPIE..627..733T, 1993ASPC...52..173T}
		}



\end{document}